\documentclass[fleqn,10pt,inline]{wlscirep}
\usepackage[T1]{fontenc}
\usepackage[utf8]{inputenc}
\usepackage[normalem]{ulem}
\usepackage{xcolor}

\usepackage[sort&compress,english,capitalise]{cleveref}         

\usepackage{orcidlink}

\usepackage{cancel}

\usepackage{siunitx}

\usepackage{makecell}

\usepackage{adjustbox}
\usepackage{amsmath}
\usepackage{mathtools}
\DeclarePairedDelimiter\abs{\lvert}{\rvert}
\makeatletter
\let\oldabs\abs
\def\abs{\@ifstar{\oldabs}{\oldabs*}}
\makeatother

\newcommand{\revtext}[1]{#1}

\newcommand{\REF}[1]{}

\newcommand{\hgi}{Hg1201}

\newcommand{\lscoi}{LSCO}
\newcommand{\thalli}{Tl2201}

\newcommand{\ptop}{$p$ to $1+p$}

\newcommand{\kBT}{\mbox{$k_\mathrm{B}T$}}
\newcommand{\ms}{\mbox{$m^*$}}

\newcommand{\muH}{\mbox{$\mu_\mathrm{H}$}}
\newcommand{\neff}{\mbox{$n_{\mathrm{eff}}$}}
\newcommand{\nH}{\mbox{$n_{\mathrm{H}}$}}

\newcommand{\nloc}{\mbox{$n_{\mathrm{loc}}$}}

\newcommand{\pst}{\mbox{$p^*$}}

\newcommand{\RH}{\mbox{$R_{\mathrm{H}}$}}

\newcommand{\Tc}{\mbox{$T_\mathrm{c}$}}

\newcommand{\Ts}{\mbox{$T^*$}}

\DeclareUnicodeCharacter{03B4}{$\delta$}
\DeclareUnicodeCharacter{2212}{-}
\DeclareUnicodeCharacter{0398}{$\Theta$}
\DeclareUnicodeCharacter{00FC}{\"u} 

\title{Transport properties and doping evolution of the Fermi surface in cuprates} 

\author[1]{B. Klebel-Knobloch\orcidlink{0000-0002-9871-4196}}
\author[2,1]{W. Tabi\'{s}\orcidlink{0000-0002-8827-9944}}
\author[1,2]{M. A. Gala\orcidlink{0000-0001-6906-9354}}
\author[3,*]{O. S. Bari\v{s}i\'c\orcidlink{0000-0002-6514-9004}}
\author[4,*]{D. K. Sunko\orcidlink{0000-0002-1383-0674}}
\author[1,4,*]{N.~Bari\v{s}i\'c\orcidlink{0000-0003-4637-0544}}

\affil[1]{Institute of Solid State Physics, TU Wien, 1040 Vienna, Austria}
\affil[2]{AGH University of Krakow, Faculty of Physics and
Applied Computer Science, 30-059 Krakow, Poland}
\affil[3]{Institute of Physics, Bijeni\v{c}ka cesta 46, HR-10000, Zagreb, Croatia}
\affil[4]{Department of Physics, Faculty of Science, University of Zagreb, Bijeni\v{c}ka cesta 32, HR-10000, Zagreb, Croatia}
\affil[*]{obarisic@ifs.hr, dks@phy.hr, nbarisic@phy.hr}


\begin{abstract}
Measured transport properties of three representative cuprates are reproduced within the paradigm of two electron subsystems, itinerant and localized. The localized subsystem evolves continuously from the Cu 3d$^9$ hole at half-filling and corresponds to the (pseudo)gapped parts of the Fermi surface. The itinerant subsystem is observed as a pure Fermi liquid (FL) with material-independent universal mobility across the doping/temperature phase diagram. The localized subsystem affects the itinerant one in our transport calculations solely by truncating the textbook FL integrals to the observed (doping- and temperature-dependent) Fermi arcs. With this extremely simple picture, we obtain the measured evolution of the resistivity and Hall coefficients in all three cases considered, including LSCO which undergoes a Lifshitz transition in the relevant doping range, a complication which turns out to be superficial. Our results imply that prior to evoking polaronic, quantum critical point, quantum dissipation, or even more exotic scenarios for the evolution of transport properties in cuprates, Fermi-surface properties must be addressed in realistic detail.
\end{abstract}

\begin{document}
\maketitle

\section*{Introduction} 

The discovery of superconductivity in 1911 was one of the most surprising in the field of solid state physics.\cite{onnes_further_1911} It took almost fifty years before the phenomenon was successfully explained by BCS theory.\cite{bardeen_theory_1957} The next milestone was the discovery of high-temperature superconductivity (SC) in cuprates about thirty-five years ago.\cite{bednorz_possible_1986} The superconducting (SC) state in these compounds is of type II, which is well understood in the BCS/London framework to mean that the coherence length is shorter than the penetration depth. In cuprates, the coherence length is extremely short, resulting in very high second critical fields, of the order of $100$~T, and the SC gap is $d$-wave, unlike elemental, phonon-mediated, superconductors where it is always $s$-wave. However, the main reason why these compounds are considered unconventional is the unusual evolution of normal-state properties with doping $p$.\cite{phillips_stranger_2022,keimer_quantum_2015} Here, one should carefully separate compound-specific from universal properties.\cite{barisic_high-tc_2022} In cuprates, SC is universally observed in the range between $p\sim 0.04$--$0.05$ (underdoped) and $0.30$--$0.35$ (overdoped), with a maximal value of the SC transition temperature (\Tc) around $p\sim 0.16$. This common pattern implies that the origin of SC stems from universal normal-state behavior, while the wide variation in observed maximal \Tc's (more than an order of magnitude) is due to more subtle non-universal effects which tune the SC in particular compounds. 

\begin{figure}%
    \centering
    \includegraphics[width=1\textwidth]{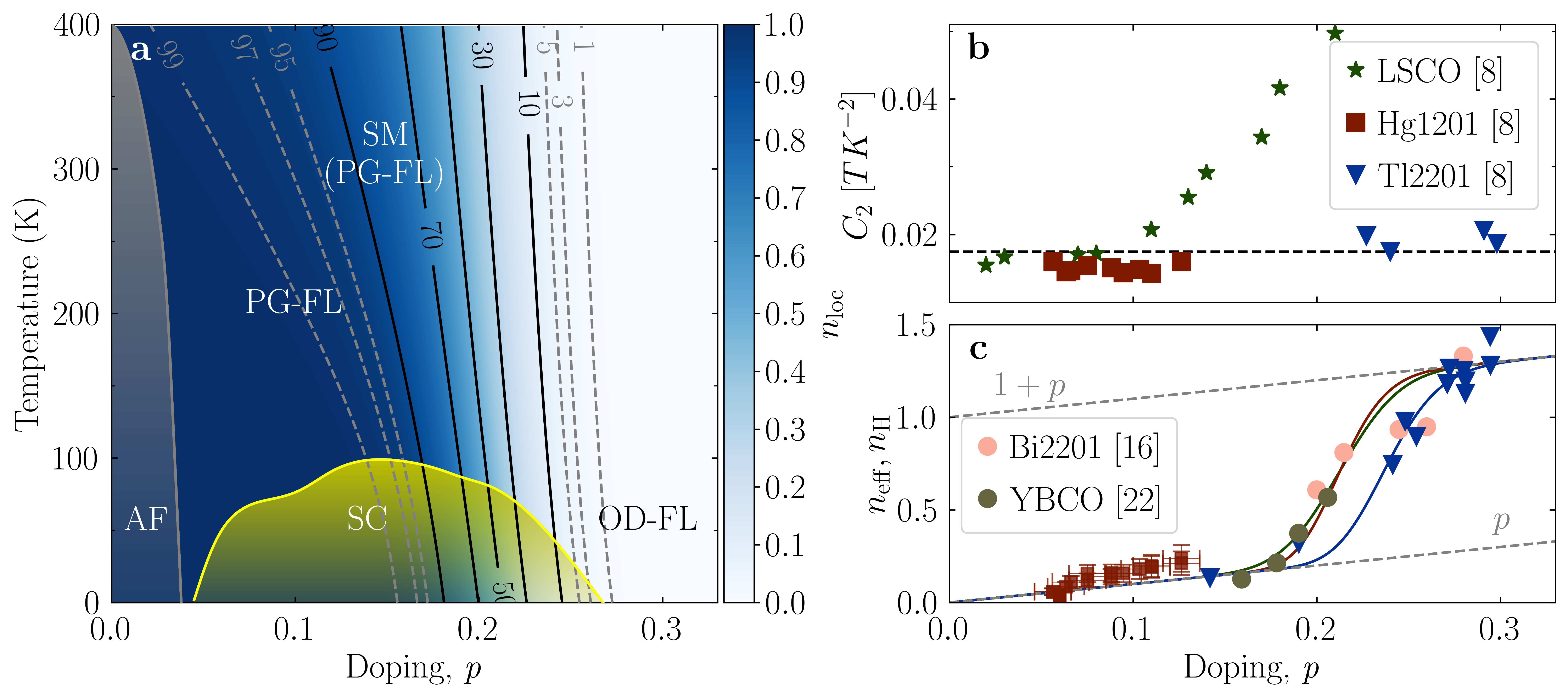}
    \caption{Phase diagram and carrier density in cuprates. \textbf{a} Schematic phase diagram that captures the evolution of key universal features of cuprates.\cite{pelc_unusual_2019, barisic_high-tc_2022} The approximate limit of the antiferromagnetic (AF) phase is shown in grey and the superconducting (SC) dome in yellow. The doping/temperature evolution of the density of localized charge, as extracted from the resistivity, is indicated by blue shading. Solid black and dashed grey lines are isodensity lines. In the OD-FL regime, all carriers contribute to electronic transport. Both, the pseudogap Fermi liquid (PG-FL, roughly corresponding to $>97$\% localized holes) and the strange metal (SM, marked by an intensive gradual delocalization) are also indicated, though there is no conceptual difference between them. \textbf{b} Measurements of the Hall mobility revealed that $C_2 =( 1/\mu_H) - C_0)/T^2$ is essentially universal in cuprates, with $C_2 = 0.0175(20)$ TK$^{-2}$ (indicated by the dashed line).\cite{barisic_evidence_2019} However, $C_2$ in \lscoi\ for $p > 0.08$ deviates strongly from the universal value. 
    \textbf{c} Doping dependence (at $T = 0$~K) of the carrier density (\neff) as determined from the resistivity (full lines),\cite{pelc_resistivity_2020} compared with values obtained from the Hall coefficient (solid points).\cite{putzke_reduced_2021,ayres_incoherent_2021} Both quantities behave identically and reveal the $p$ to $1 + p$ change in the carrier density.
        }\label{fig:intro_C2}
\end{figure}

Indeed, despite many compound-specific properties within this group of materials, a range of underlying universal behaviors was identified\cite{barisic_high-tc_2022} precisely in those normal-state transport properties which were long considered to be both the key to their SC and widely, but wrongly, taken as proof that the charge carriers were not a Fermi liquid (FL).\cite{phillips_stranger_2022,keimer_quantum_2015} A particular milestone in establishing that the itinerant carriers were, in fact, a FL was the observation that the sheet resistance (i.e., resistance per CuO$_2$ layer) in these compounds is universal.\cite{barisic_universal_2013} But perhaps the most surprising universality is that the Hall mobility across the doping-temperature phase diagram of the cuprates is essentially compound- and doping-independent, as discovered through combined measurements of the resistivity ($\rho$) and Hall coefficient (\RH).\cite{barisic_evidence_2019} Moreover, it was shown that the Hall mobility ($\mu_H^{-1} =\frac{\rho}{\RH}=\frac{m^{\ast}}{e\tau}$) exhibits a robust quadratic temperature dependence ($\mu_H^{-1} = C_2 T^2$), with an essentially universal value of $C_2=$~\SI{0.0175(20)}{\tesla\kelvin^{-2}}, as presented for Hg1201, Tl2201 and LSCO at low doping ($p < 0.08$) in \cref{fig:intro_C2}b (for other cuprate compounds see Ref. \citeonline{barisic_universal_2013}).
    
The universal quadratic dependence of $\mu_H^{-1}$ suggests that the underlying transport scattering rate is FL-like in all temperature and doping regimes of the relevant phase diagram.\cite{barisic_evidence_2019} And indeed, the FL nature of itinerant charges was unambiguously demonstrated in both regimes (under- and overdoped) by experimental observations, e.g., FL scalings in the underdoped regime,\cite{mirzaei_spectroscopic_2013,kumar_characterization_2023,chan_-plane_2014} or the Wiedemann-Franz law,\cite{proust_heat_2002} angle-resolved photoemission spectroscopy (ARPES),\cite{plate_fermi_2005} and quantum oscillation measurements in the overdoped regime.\cite{vignolle_quantum_2008} These fundamental experimental facts imply that the explanation for the behavior of itinerant charges in cuprates must be searched for first within the standard framework of FL charge transport.

The universality of the Hall mobility implies a well-defined, fixed ratio $\frac{m^*}{e\tau}$. Consequently, the resistivity (${\rho}={R_H}\mu_H^{-1}$) provides direct information about the carrier density in the FL framework.\cite{barisic_universal_2013} A systematic analysis of the extensive electronic transport data allows one to determine how the effective carrier density \neff\ evolves across the temperature-doping phase diagram. This analysis, summarized in \cref{fig:intro_C2}a,c, reveals that in the low-temperature limit, the effective \emph{itinerant} carrier density \neff\ changes gradually with decreasing doping from $\neff = 1 + p$ to $\neff = p$.\cite{pelc_emergence_2018,barisic_evidence_2019} Denoting the density of \emph{localized} carriers by \nloc, the total carrier density satisfies the relation
\begin{equation}
\nloc + \neff = 1 + p
\label{eq:charcons}
\end{equation}
by charge conservation. Hence, the change in \neff\ means that exactly one hole carrier per CuO$_2$ unit cell localizes ($\nloc: 0 \rightarrow 1$) when crossing from the overdoped to the underdoped region of the phase diagram. Such an evolution of the effective carrier density extracted from resistivity measurements was confirmed several years later by measurements of the doping evolution of the high-field low-temperature Hall number $n_H = \frac{V}{eR_H}$ ($V$ is the elementary cell volume) determined in Bi2201 and Tl2201, shown in \cref{fig:intro_C2}c (for YBCO a correction for the anisotropy factor was required\cite{putzke_reduced_2021}). It was also demonstrated, by transport\cite{barisic_evidence_2019} and optical conductivity,\cite{kumar_characterization_2023} that the incipient change in the effective carrier density just below optimal doping is also responsible for the linear-in-temperature resistivity observed in this so-called ``strange metal'' (SM) regime (\cref{fig:intro_C2}a). Consequently, some of us attributed the whole unusual evolution of different properties in cuprates to the localized charge, in particular to its gradual delocalization with temperature or doping.\cite{barisic_evidence_2019,barisic_high-tc_2022} Notably, that is by definition the non-Fermi-liquid component of the cuprate problem, because localized charges do not conduct.

Currently, countless alternative interpretations of normal-state properties are based solely on the apparent non-Fermi liquid evolution of the scattering rate, focusing exclusively on the optimally doped or overdoped regimes, and without taking into account that the carrier concentration (i.e., the density of states at the Fermi level) can also change. For example, it is often argued that the linear temperature dependence of resistivity is caused by underlying quantum criticality,\cite{cooper_anomalous_2009} or, according to most recent interpretations, by the charge scattering rate reaching\cite{legros_universal_2019} the Planckian limit,\cite{Zaanen04, sadovskii_planckian_2021} where it is further argued that this scattering is momentum-independent and inelastic.\cite{grissonnanche_linear-temperature_2021} Furthermore, the significant reduction in the Hall number from $1+p$ to $p$ was attributed to quasiparticle decoherence, despite the fact that the determined \nH\cite{putzke_reduced_2021} perfectly coincides with \neff\ determined earlier from the resistivity.\cite{pelc_emergence_2018} It was recently suggested that cuprates are best understood in terms of two distinct current-carrying fluids, of which one behaves like a coherent FL.\cite{ayres_incoherent_2021,ayres_superfluid_2022} Thus, even within the scattering-rate scenarios alone, the electronic properties of cuprates are intensely discussed, with mutually incompatible proposals.\cite{phillips_stranger_2022}

A recent analysis of the optical conductivity data clearly separates scattering-rate from carrier-density effects, revealing unequivocally that the missing part of the Fermi surface (FS) outside the well-known arcs is indeed gapped in cuprates.\cite{kumar_characterization_2023} In the present work, we follow the same gapping scenario to calculate the values of $R_H$ and the longitudinal conductivity $\sigma_{xx}$ ($=1/\rho$) as a function of doping directly from the measured FS, and compare them with experimental data. The ungapped segments, Fermi arcs, are the only parts of the FS that contribute to charge transport. After establishing the calculation procedure on compounds with rather simple, nearly circular underlying FS's, we focus specifically on \lscoi. It undergoes a Lifshitz transition \revtext{(change from hole- to electron-like topology)} in the doping range of interest,\cite{yoshida_systematic_2006} which presents a challenge to our simple FS approach.\cite{barisic_universal_2013} Indeed, as the Lifshitz transition is approached in doping, the values of $C_2$ and \nH\ in \lscoi\ strongly deviate from their universal values, both quantitatively and qualitatively, as shown in \cref{fig:intro_C2} b and c, respectively. We show that even such strong deviations are captured in considerable detail by the suggested (universal FL) calculation approach. Thus, the exception of LSCO turns out to be superficial. Rather, it serves only to corroborate the universality. Because the appearance of the Lifshitz transition in parallel with the gradual (de-)localization process explains even strong deviations from universal behaviors, we will argue that prior to applying exotic approaches to analyze any particular compound, one should try to carefully establish the exact shape of the FS first, and check if the same, quite standard, procedure can be applied. Finally, because our calculations unambiguously show that the whole complexity of cuprates stems from the gradual localization of exactly one charge per CuO$_2$ plaquette, its role in the superconducting mechanism will be discussed as well.

\section*{Results} 
        
We begin our analysis by invoking the standard definition of the Hall coefficient $R_H$ in terms of the directly measurable diagonal ($\sigma_{xx}, \sigma_{yy}$) and off-diagonal ($\sigma_{xy}$) components of the conductivity tensor:\cite{ong_geometric_1991,niksic_multiband_2014,Kupcic17}
\begin{align}
R_H &= - B^{-1} \frac{\sigma_{xy}}{\sigma_{xx}\sigma_{yy}},  \label{eq:R_H} 
\end{align}
with B the applied magnetic induction. For the tensor terms, we use standard FL expressions:\cite{ong_geometric_1991,niksic_multiband_2014,Kupcic17,culo_possible_2021}
\begin{align}
    \sigma_{xx}&= \frac{e^2\tau}
    {2\hbar^2}\frac{N_V}{\Gamma_{2D}}\oint_{E_F}dk_\parallel\;\abs{\frac{\partial\varepsilon_{\vec k}}{\partial k_\perp}}
    \label{eq:sigma_xx} \\ 
    \sigma_{xy}&=\frac{e^3B\tau^2}{\hbar^4}\frac{N_V}{\Gamma_{2D}}\oint_{E_F} dk_\parallel\;\abs{\frac{\partial\varepsilon_{\vec k}}{\partial k_\perp}}^{-1}
    \left[\left(\frac{\partial\varepsilon_{\vec k}}{\partial k_x}\right)^2
    \frac{\partial^2 \varepsilon_{\vec k}}{\partial^2 k_y}
    -\frac{\partial\varepsilon_{\vec k}}{\partial k_x}
    \frac{\partial\varepsilon_{\vec k}}{\partial k_y}
    \frac{\partial^2 \varepsilon_{\vec k}}{\partial k_y\partial k_x}\right]\;.
    \label{eq:sigma_xy}
\end{align}

\noindent Here, $k_\perp$ and $k_\|$ are components of the charge carrier wave vector ${\vec k}$ perpendicular and parallel to the FS, respectively. $\Gamma_{2D}$ is the area of the 2D Brillouin zone and $N_V$ is the number of states per unit volume. The integrals are usually taken over the whole FS. However, because parts of the FS are gapped in cuprates, only the Fermi arcs centered at the nodes contribute to these integrals. To describe the arc lengthening with doping, we introduce a parameter $f_g = \neff/\left(1+p\right)$, where the evolution of \neff\ is inferred from resistivity measurements.\cite{pelc_unusual_2019} Quantitatively, $f_g$ denotes the fraction of ungapped states contributing to the transport on the FS, relative to the full underlying FS. The doping evolution of $f_g$ is presented in \cref{fig:summary_Hg1201Tl2201}b and \cref{fig:summary_LSCO}b for our three representative materials. Hereafter, the integrations in \cref{eq:sigma_xx,eq:sigma_xy} are understood to be carried out only along the Fermi arcs, whose length is expressed by $f_g$. 

It can be seen without calculation that an ideal partially gapped parabolic band (i.e., circular underlying FS) immediately leads to a 1:1 correspondence between \neff\ and \nH. Simply, because the Fermi velocity $v_F=\hbar^{-1}\abs{\partial\varepsilon(\vec k_F)/\partial k_\perp}$ and scattering rate do not change along the FS, the value of the integrals must give the length of the arc $f_g$, which in turn directly corresponds to \neff{} (for a more detailed discussion see Methods, \cref{sec:parabolic_band}).

However, the fact is that the underlying FS is non-universal for different cuprates, exhibiting curvatures with a significant departure from the circular form, with different values of Fermi velocities along the FS. Thus, to calculate  $\sigma_{xx}$ and $\sigma_{xy}$ from \cref{eq:sigma_xx,eq:sigma_xy}, respectively, knowledge of the exact shape of the bands in the \kBT\ window around the Fermi level (i.e., the only energy range relevant for transport) is required. ARPES experiments measure the dispersion of bands near the Fermi energy directly. In the case of single-layer cuprates, only one band intersects the Fermi level. This band may be parametrized by 2D tight-binding models (for more details see Methods \cref{sec:app:TB-details}). Here, we have used previously published best-fit parametrizations of ARPES data, essentially without any further modification.

\subsection*{Doping evolution of $R_H$ in Hg1201 and Tl2201: the case of nearly circular FS's}

        \begin{figure}%
            \centering
            \includegraphics[width=0.9\textwidth]{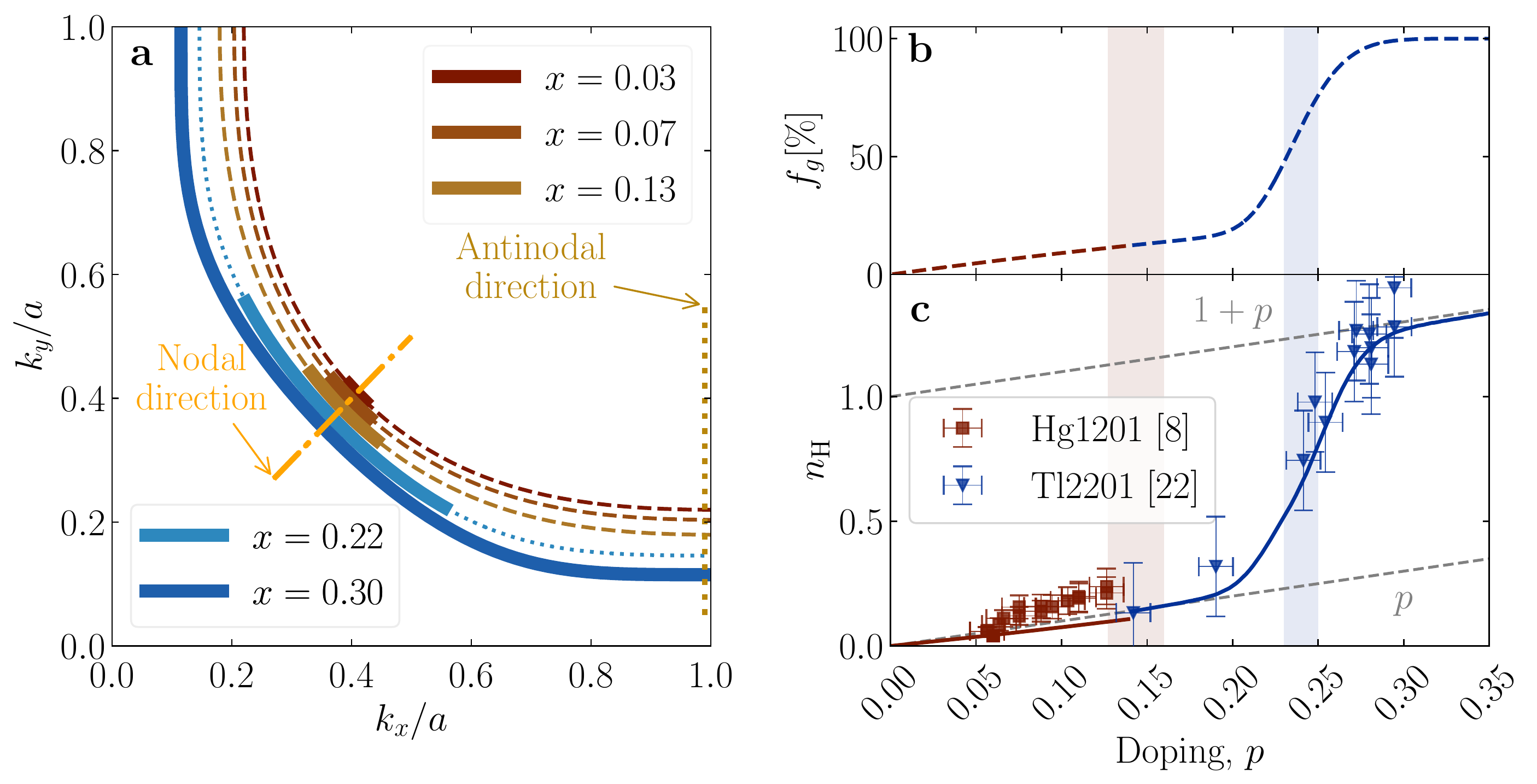}
            \caption{
            The FS and Hall number of Hg1201 and Tl2201. 
            In \textbf{a}, the FS's as parametrized in Refs.~\citeonline{das_q0_2012,vishik_angle-resolved_2014} (Hg1201) and \citeonline{plate_fermi_2005,peets_tl2ba2cuo6_2007} (Tl2201) are shown. The underlying FS's are almost circular as obvious from the dashed (Hg1201) and dotted (Tl2201) lines, where arcs (ungapped states) are indicated with full lines.
            \textbf{b} The arc-length, $f_g$, as extracted from the resistivity. 
           \textbf{c} The here calculated \nH\ (full line) is compared with the measured values (points) from Refs.~\citeonline{barisic_evidence_2019} (\hgi{}) and \citeonline{ayres_incoherent_2021} (\thalli{}), where error bars are reproduced from the respective cited works. 
            The shaded areas in \textbf{b} and \textbf{c} indicate the doping ranges for which ARPES data is available. In the case of \hgi{}, experimental values of \nH{} are collected at $T=$ \SI{100}{\kelvin}, just above the value of the maximal \Tc{} ($\sim 95$ K) in this compound, while for \thalli{} high-field zero-Kelvin extrapolations are shown. 
            }\label{fig:summary_Hg1201Tl2201}
        \end{figure}
        
The underlying FS's of \hgi{} and \thalli{} are nearly circular. We also recall that \muH\ in both compounds practically does not change with doping or temperature as the arcs lengthen, which implies that all (arc) segments have the same or nearly the same contribution in terms of $v_F$'s and scattering rates. Thus, it is expected that the calculated \nH\ correctly corresponds to \neff{} (determined from resistivity), if one chooses $f_g = \neff/\left(1+p\right)$ for the range of the integrals in \cref{eq:sigma_xx,eq:sigma_xy}.

Fermi surfaces were previously measured by ARPES and parametrized by effective tight-binding models at $p\sim 0.15$ for \hgi{} \cite{das_q0_2012,vishik_angle-resolved_2014} and at $p\sim 0.24$ for \thalli{} \cite{plate_fermi_2005,peets_tl2ba2cuo6_2007} (see \cref{tab:TB-parameters} in the Methods section). These doping levels are indicated by shaded vertical bands in \cref{fig:summary_Hg1201Tl2201}, right panel. To extend our calculations to other doping levels of interest without introducing fitting ambiguities, we shifted the chemical potential in the same (rigid) band to satisfy Luttinger's sum rule for the underlying, ungapped FS (see \cref{eq:rigid_band_shift} in the Methods section). Even such a crude, zeroth-order approximation turns out to be sufficient to correctly capture the doping evolution of transport coefficients in \hgi{} and \thalli{}. To determine \neff{} from the resistivity in the crossover region of \ptop{}, the approach introduced in Ref.~\citeonline{pelc_unusual_2019} was used. Again, only previously published parameters were used (Ref. \citeonline{pelc_unusual_2019} for \hgi{} and Ref. \citeonline{pelc_resistivity_2020} for \thalli{}), with the new parameter $f_g$ fixed by transport (with details in the Methods section). The resulting FS's are shown in  \cref{fig:summary_Hg1201Tl2201}. By the dashed (dotted) lines we indicate the underlying FS of \hgi{} (\thalli{}), while full lines indicate the length of the ungapped arcs at specific doping levels, which correspond to the fractional extension of the FS shown in \cref{fig:summary_Hg1201Tl2201}b. The nearly circular shape of the underlying FS's is apparent. From the integration over the arcs, we calculate the Hall number according to \cref{eq:R_H} (full line in \cref{fig:summary_Hg1201Tl2201}c), and compare the result with measured values of \nH\ from \hgi{} (red) and \thalli{} (blue). Unsurprisingly, the calculated doping dependence of \nH\ (and $\rho$ discussed below) in the limit of $T=0$ correctly represents \neff{}, as shown in \cref{fig:intro_C2}c. As already noted earlier,\cite{barisic_evidence_2019} the measured \nH\ of \hgi{} (see \cref{fig:summary_Hg1201Tl2201}c,) is slightly higher than the calculated values, due to subtle difficulties in determining the exact sample geometry as well as the concentration of holes in the CuO$_2$ layer accurately in cases of interstitial oxygen doping.

\subsection*{The case of a Lifshitz transition in LSCO: the exception that confirms the rule }
        
        \begin{figure}%
            \centering
            \includegraphics[width=0.9\textwidth]{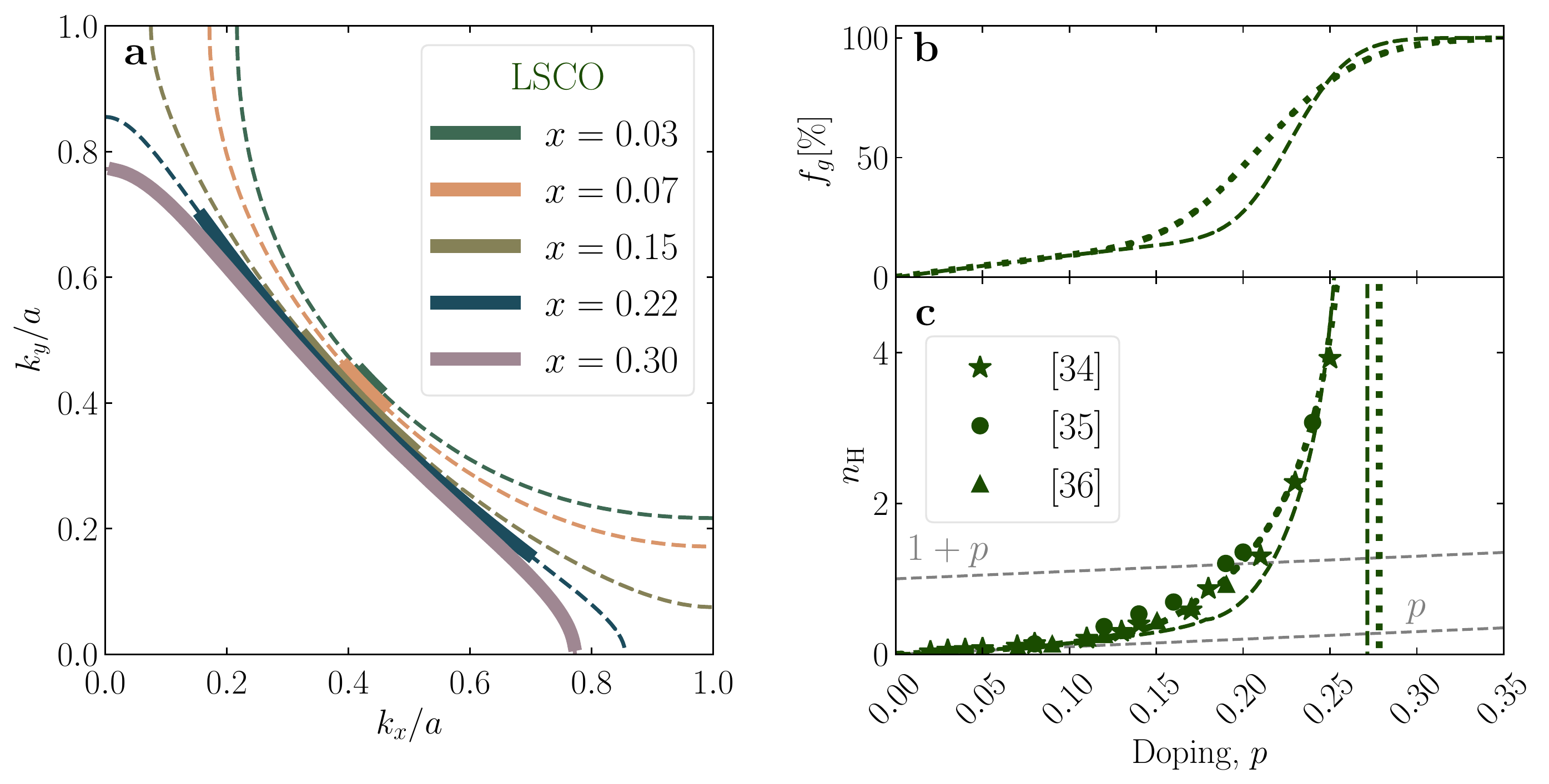}
            \caption{
            The FS and Hall number of LSCO. 
            In \textbf{a}, the FS as parametrized in Ref.~\citeonline{yoshida_systematic_2006} is shown, where the dashed lines correspond to the underlying FS, while the arcs appear as full lines. The underlying FS undergoes a Lifshitz transition between  $0.15<p<0.22$. \textbf{b} The arc length, $f_g$, as determined from the resistivity is shown as dashed line. In case of LSCO, $f_g$ was also adjusted to obtain a better fit of the Hall data. The resulting evolution is shown by the dotted line. \textbf{c} The combination of the here calculated \nH{} (dashed and dotted lines) and previously reported experimental data (Refs.~\citeonline{ando_evolution_2004,tsukada_negative_2006,padilla_constant_2005}) reveals an excellent agreement. 
            Consistently with the case of \hgi{} but also to avoid problems related the ordering tendencies in LSCO at low temperatures,\cite{li_hidden_2016,barisic_universal_2013} the experimental values of \nH\ are collected at $T=$\SI{100}{\kelvin}. Higher doping levels and the temperature dependence are discussed in the Supplementary Information 1. 
            }\label{fig:summary_LSCO}
        \end{figure}

To put our approach to a more challenging test, we extend the same analysis to LSCO, whose FS has a rather interesting evolution with profound consequences on transport coefficients. In this compound, a Lifshitz transition in the crossover region of \ptop\ is well established. Notably, a similar Lifshitz transition has been seen by ARPES in the bismuth cuprates (Bi,Pb)$_2$(Sr,La)$_2$CuO$_{6+\delta}$ and Bi$_2$Sr$_2$CaCu$_2$O$_8+\delta$ as well.\cite{kondo_hole-concentration_2004,piriou_first_2011,kaminski_change_2006,drozdov_phase_2018} However, the Lifshitz transition occurs at higher doping levels there (in the single-layer compound, at $p\gtrsim 0.3$), at the limit of synthesis capabilities. Therefore, it is both less interesting and less convenient to investigate the Lifshitz transition in bismuth cuprates. On the other hand, ARPES measurements of \lscoi\ have been extensively documented for a wide doping range, thus the FS is established exceptionally well. Moreover, the band-structure of \lscoi\ was parameterized through tight-binding parameters as reported in Refs.~\citeonline{yoshida_low-energy_2007,yoshida_systematic_2006}. This parametrization includes doping-dependent tight-binding parameters (see Methods). Notably, with this published parametrization, the total carrier density of the underlying FS deviates slightly from Luttinger's sum rule. However, this roughness introduces only a small uncertainty in our calculations, which is henceforth neglected, highlighting the underlying stability of our approach. To extrapolate between measured doping levels, the tight-binding parameters were interpolated by smooth polynomials (see Methods \cref{tab:TB-parameters}). 
        
In \cref{fig:summary_LSCO}a, we show the FS of \lscoi\ parametrized according to Ref.~\citeonline{yoshida_systematic_2006}. It accurately reproduces the ARPES measurement, in particular the change from the hole-like circular shape to an electron-like diamond shape with increasing $p$.\cite{yoshida_low-energy_2007} To calculate $\sigma_{xx}$ and $\sigma_{xy}$, we follow exactly the same procedure as above for \hgi{} and \thalli{}. The doping evolution of \neff{} is determined from the resistivity,\cite{pelc_unusual_2019} which in turn defines the length of the arcs $f_g$. The evolution of the underlying FS and the concomitant change of the arc length is displayed in \cref{fig:summary_LSCO}, a and b, respectively. Taking into account the simplicity of our approach against the complexity of the underlying FS, the calculated \nH\ agrees surprisingly well with measured values, as shown in \cref{fig:summary_LSCO}c and the Supplementary Information, Figures S1 and S2.

We have thus obtained a simple understanding of why the large deviation of \nH\ from \neff{} in \lscoi, overshooting  $1 + p$ divergently, does not invalidate our general FL approach for the arc carriers in cuprates. Namely, the anomaly is a direct manifestation of the Lifshitz transition in the underlying FS, which causes the \emph{denominator} in the FL expression for \nH, which measures FS curvature, to go through zero as the FS changes from hole- to electron-like.\cite{niksic_multiband_2014,Kupcic17} Concurring with that interpretation, negative values of \nH\ have been reported in thin films at $p \geq 0.32$.\cite{tsukada_negative_2006,yoshida_low-energy_2007} This re-entrance of negative \nH\ values with doping emerges naturally from our analysis, as further discussed in Supplementary Information 1.

Parenthetically, we mention that, because of the complex shape of the FS and the changes of $f_g$ with doping, it is inherently difficult to define at which exact doping the Lifshitz transition is supposed to occur. If we define this transition as the point where the underlying FS curvature changes sign from hole-like to electron-like, we can pinpoint it at $p\sim0.18$. However, because parts of the FS are gapped at this doping, this point is barely noticeable, as a minuscule kink in \cref{fig:summary_LSCO}c (hidden by a measured point), and a kink in \cref{fig:res_tauA2sheet}a. On the other hand, if we consider the doping dependence of \nH\ as primary, and interpret its point of divergence as the Lifshitz transition, this puts it at a significantly higher doping level of $p\sim 0.28$. This difference shows that the precise position of the Lishitz transition in cuprates manifests itself differently in \nH\ and in dispersions fitted to ARPES. Finally, we note that \nH\ starts to deviate from \neff{} even below the \ptop{} crossover. This happens because the arcs begin to flatten due to the proximity of the Lifshitz transition, with the flat sections making a small contribution to the curvature. Importantly, it follows from the same reasoning that the divergence in \nH\ cannot affect the $p$ to $1+p$ crossover in \neff, because the latter is measured simply as the total number of itinerant carriers, irrespective of the shape of the FS.

\subsection*{Resistivity}
    
        \begin{figure}%
        \centering
        \includegraphics[width=1\textwidth]{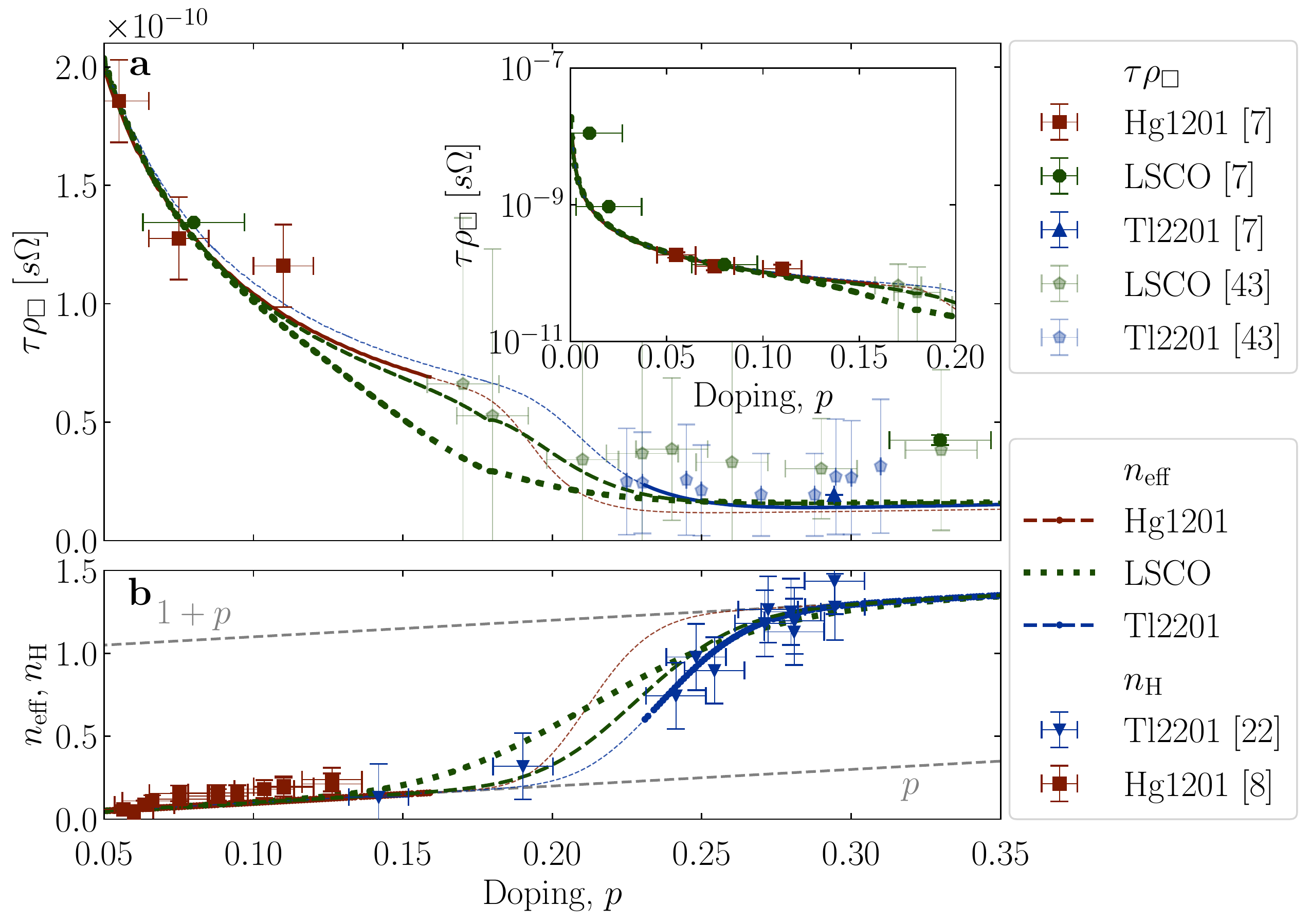}
        \caption{
        Calculated \textit{versus} measured resistivity and carrier density. \textbf{a} To facilitate the comparison between calculated (lines) and measured sheet resistance ($\rho_{\Box}\sim A_2T^2$ -- (opaque symbols)\cite{barisic_universal_2013} or in the crossover regime $\rho_{\Box}\sim A_1T^1 + A_2T^2$ -- (shaded symbols)\cite{hussey_generic_2013}) a temperature independent quantity $\tau\rho_{\Box} \sim A_2/C_2$ is displayed, as a function of doping, for all three discussed compounds (see the Methods \cref{sec:methods:resistivity} for details). Dashed and dotted lines for \lscoi{} correspond to calculations with the same $f_g$ as presented by dashed and dotted lines in \cref{fig:summary_LSCO}b. 
        The inset shows an extended doping range to $p=0$ on a logarithmic scale for clarity. A small kink in the calculated doping dependence for LSCO at $p\sim 0.18$ (dashed line) coincides with the Lifshitz-transition of the underlying FS. 
        \textbf{b} Full and dashed lines show \neff{} as inferred from resistivity measurements, which is the only input parameter for the performed calculation. In case of LSCO, an additional dotted line indicates \neff{} obtained by adjusting $f_g$ (i.e., arc-length) for a  better fit of the Hall data. For Hg1201 and Tl2201 \neff{} (lines) and \nH\ (symbols) coincide. This is not the case for LSCO, where \nH\ diverges at the Lifshitz transition. However, \neff{} shows a similarly smooth crossover in LSCO as it does in Hg1201 and Tl2201.
        The calculated $\sigma_{xx}$ [\cref{eq:sigma_xx}] strongly depends on $v_F$, whose value is usually not controlled in tight-binding fits to ARPES data. Therefore, normalization factors $f_{\mathrm{norm}}$ have been applied to $\tau\rho_{\Box}$ of \lscoi\ and \thalli{}. The details of this normalization are in Supplementary Information 2. 
        }
        \label{fig:res_tauA2sheet}
    \end{figure} 

Now we turn back to the resistivity to demonstrate the robustness and self-consistency of the above analysis. It is also a necessary step because we have originally relied only on the universality of \muH\ to determine \neff{} and consequently the length of the arcs, neglecting all deviations, including the (large) one shown in \cref{fig:intro_C2} for \lscoi. In this determination of \neff, all of the underlying $v_F$'s were tacitly taken to be universal because the mobility is essentially universal. Now, we will calculate the doping dependence of the resistivity from the arced FS's, using \cref{eq:sigma_xx}, which takes into account the variation of $v_F$ along the arc, but strictly respecting the experimentally established universality of the nodal $v_F$ (see also Supplementary Information 2 for details).\cite{zhou_high-temperature_2003} To compare our calculations with experimentally established values, we plot the results in the form of $\tau\rho_{\Box}=A_2/C_2$, with $A_{2\Box}$ as in $\rho_{\Box}\sim A_{2\Box}T^2$ from Ref.~\citeonline{barisic_universal_2013} combined with $C_2$ as in $\tau^{-1}\sim C_2T^2$ from Ref.~\citeonline{barisic_evidence_2019}. $A_2$ and $C_2$ are both pre-factors to a squared temperature behavior, so the temperature cancels in the product, i.e., $\tau\rho_{\Box}=A_2/C_2$ is a temperature-independent parameter (see the details in the Methods section). In this way, we can compare data measured at finite temperatures with our calculation at $T=0$. As obvious from \cref{fig:res_tauA2sheet}, the agreement is remarkable, which is perhaps expected in case of \hgi{} and \thalli{} but less so, given the simplicity of our approach, in case of \lscoi{}. This agreement also implies that $v_F$ does not vary significantly along the parts of the arcs with a significant contribution to transport.

Finally, to test our approach even further, we invert it and fit our calculation to the measured \nH, to obtain \neff{} which defines the arc length $f_g$ (See \cref{tab:gap-parameters} in the Methods for details). In this case, the agreement between measured and calculated values of \nH\ is by design (dotted line in \cref{fig:summary_LSCO}c). It might be interesting to note that this approach results in a somewhat broader \ptop{} crossover than reported elsewhere,\cite{ayres_incoherent_2021} as shown in Figs. 3b and 4b. This is perhaps to be expected since \lscoi\ is a compound that is known to be disordered. However, the two approaches are qualitatively the same and the (rather small) difference sets the limits of the expected uncertainty.

The overall universality of the sheet resistance can be understood given that the Cu $3d$ orbital is blocked by Coulomb effects, so coherent FL conduction dominantly occurs via the Cu $4s$--O $2p_{x,y}$, and secondarily via the O $2p_x$--$2p_y$, orbital overlaps.\cite{barisic_high-tc_2022} Both are chemically invariant across the cuprates\cite{Pavarini01} in agreement with the universality of $v_F$ along the arcs established here. Notably, in LSCO at the antinodes, $v_F$ has a strong doping dependence due to the Lifshitz transition. However, as apparent from the above, these parts of the FS contribute to transport processes only when they become ungapped at elevated doping levels, at which point the van Hove singularity (vHS) has moved away from the FS again. Therefore, the sheet resistance satisfies the universal value in a broader doping range than would be expected from considering $v_F$ along the full underlying FS. 
    
\section*{Discussion}

In the context of the last 35 years of debates in the field of cuprates, each new demonstration that textbook FL formulas can be used to describe a key property of these materials marks essential progress. Here, we have shown that these formulas are perfectly adequate to calculate transport coefficients, even in the complex case of a concurrent Lifshitz transition, while it has been shown elsewhere that they can also describe other key properties, like optical conductivity,\cite{kumar_characterization_2023,mirzaei_spectroscopic_2013} specific heat\cite{zhong_differentiated_2022}, magneto-resistivity\cite{chan_-plane_2014}, quantum oscillations\cite{vignolle_quantum_2008,doiron-leyraud_quantum_2007,barisic_universal_2013-1,tabis_arc--pocket_2021}, etc. This robustness implies that, even when large deviations from the reported universal behaviors in cuprates are observed, one should seek first to understand them by taking the actual shape of the FS carefully into account\cite{barisic_evidence_2019,culo_possible_2021}, distinguishing between the localized and the itinerant charges.

Fermi arcs in cuprates have been extensively discussed, mostly from the point of view of intraorbital interactions (large Hubbard $U_d$) and the concomitant AF correlations, which were associated with the pseudogap. Such approaches have difficulties with the proper estimation of the amount of mobile charge available to conduction, and to the Hall effect in particular. To our knowledge, L.~Gor'kov and G.~Teitel'baum were the first to estimate the FL carrier concentration from the length of the arcs relative to the total underlying FS,\cite{Gorkov14} as we have done here. The observation that \nH\ diverges in \lscoi\ because of the Lifshitz transition has been made previously by I.~Kup\v{c}i\'{c} and S.~Bari\v{s}i\'{c}.\cite{niksic_multiband_2014,Kupcic17} Here, we have harnessed this phenomenology to answer a precise question: Can the deviation of the quadratic temperature coefficient $C_2$ of the Hall mobility in \lscoi\ from its universal constant value in all other cuprates be wholly explained within the same simplest-possible FL framework? The answer is yes: Once the carrier concentration is read off from the arc lengths, and the Lifshitz transition is taken into account, there is nothing specific left to model in \lscoi.

While the narrow point so made is impressive enough---there is really no exception to the universal properties of the conducting FL in cuprates---its indirect repercussions are even greater. It means that the material-specific properties, among which the value of \Tc\ is the most significant, are entirely regulated by the other component in the charge-conservation equation, \cref{eq:charcons}, namely the localized hole. It confirms that the pseudogap itself is a signature of that hole localization, not of the interactions among itinerant carriers in the arc. The latter was the default assumption of many previous investigations, including the ones cited above.

This interpretation of the pseudogap is expected both on later theoretical and independent experimental grounds. In the meanwhile, Fermi arcs have been obtained in a one-body DFT+U calculation,\cite{Lazic15} once the Coulomb doping mechanism\cite{Mazumdar89} has been correctly taken into account, with its concomitant in-plane orbital disorder. Experimentally, optical spectroscopy shows the localized hole as a clearly gapped mid-infrared feature, once the FL signal calculated from transport is subtracted.\cite{kumar_characterization_2023} These investigations and the present one concur that there are really no itinerant states at the Fermi energy beyond the arcs, so there is no need for any special mechanism---quantum dissipation, or pocket reconstruction, to name but a couple of more popular scenarios---to account for their absence in ARPES. All that is needed is to acknowledge that the pseudogap originates physically in the background (ionic) Coulomb forces which localize part of the charge, not in the interactions between the itinerant carriers. 

A number of observations with putative quantum-critical-point interpretations have turned out to be something else on closer inspection. For example, it was recently reported in Ref.~\citeonline{zhong_differentiated_2022} that the maximum in the electronic specific heat found\cite{momono_low-temperature_1994,girod_normal_2021} around $p\sim 0.20-0.22$ can be related to a Lifshitz transition by standard expressions for the electronic specific heat based on a tight-binding parametrization of ARPES data. This approach is similar to ours, and with the same conclusion, that it is not necessary to introduce a quantum critical point \pst\ at $p=0.19$ to reconcile calculations with the data.

The apparent discontinuity in the evolution of \nH\ with doping in YBCO was also originally claimed to imply a quantum critical point,\cite{badoux_change_2016} despite the fact that \neff{} estimated from resistivity, reported earlier, showed a gradual \ptop{} evolution.\cite{pelc_unusual_2019} However, this discontinuity disappeared when the chain anisotropy was taken into account,\cite{putzke_reduced_2021} as already mentioned in the Introduction.

In order to apply the above scheme effectively, some simple pitfalls should be avoided. First, different probes will sometimes see different arc lengths, or a Lifshitz transition at (slightly) different doping levels. Here, the key is that the orbital transition by which the hole delocalizes can be triggered by the probe itself, most easily by temperature, so one observes a considerable change in \neff{} as the temperature rises above \Ts,\cite{kumar_characterization_2023,barisic_evidence_2019} resulting in elongation of the arcs in a similar manner as demonstrated here as a function of doping. In the context of the Lifshitz transition, the impression will be that it is approached at lower doping levels with a higher-energy probe smeared with its accompanying finite width. In particular, as shown here, it is seen in ARPES sooner than in transport.\cite{zhong_differentiated_2022} Second, and more importantly, one should distinguish dispersive and diffusive conduction. If the same (coherent) carriers (quasi-particles) encounter several scattering mechanisms, say internal (umklapp) scattering and impurities, these will add to the total \emph{resistivity}:
\begin{equation}
\frac{1}{\tau_{\mathrm{tot}}} = \frac{1}{\tau_{\mathrm{int}}}
+ \frac{1}{\tau_{\mathrm{imp}}}.
\label{mathiessrule}
\end{equation}
On the other hand, if a part of the dispersive carriers becomes diffusive for an unspecified reason so that there are two conductive subsystems at the same time, their contributions will add to the total \emph{conductivity}:
\begin{align}
\sigma_{\mathrm{tot}} &= \sigma_{\mathrm{coh}} + \sigma_{\mathrm{diff}} \\
&= \frac{e^2}{\ms} \left(  n_{\mathrm{coh}}\tau_{\mathrm{coh}}
+ n_{\mathrm{diff}}\tau_{\mathrm{diff}} \right).
\end{align}
Assuming that the coherent part is due to internal FL scattering, $\tau_{\mathrm{coh}}=\tau_{\mathrm{int}}\sim T^{-2}$, and the temperature dependence of $\sigma_{\mathrm{diff}}$ may be anything but $T^{-2}$, one finds
\begin{equation}
\rho_{tot} = \frac{1}{\sigma_{tot}} \sim \frac{1}{T^{-2} + \sigma_{\mathrm{diff}}} \sim T^\alpha.
\end{equation}
In other words, the pure FL $T^2$ behavior is contaminated by the diffusive component, resulting in an effective power law with a real-number exponent $\alpha\neq 2$. The experimental fact that we can detect a clean $T^2$ behavior of the \emph{resistivity} deep in the PG regime, at low temperatures,\cite{barisic_universal_2013,tabis_arc--pocket_2021} close to T$_c$,\cite{popcevic_percolative_2018} shows that any contribution of putative incoherent carriers is completely negligible. Furthermore, irradiation of the sample produces simple offsets of the origin of the $T^2$ according to Mathiessen's rule,\cite{rullier-albenque_disorder_2008} \cref{mathiessrule}, essentially ruling out all but FL explanations. Notably, the same conclusion can be drawn from the Hall mobility. Not only is this property ($C_2T^2$) universal across the phase diagram,\cite{barisic_evidence_2019} but also the constant term related to impurity scattering ($C_0$) was documented very early, precisely in the so-called strange metal regime at optimal doping.\cite{chien_effect_1991}

Interestingly, the incoherent contribution is observed in pnictides,\cite{kumar_characterization_2023} where a vHS at the Fermi level\cite{derondeau_fermi_2017} provides a ready reservoir of slow carriers, easily turned diffusive even by the lowest temperatures. Such a contribution should also be expected in cuprates in which the vHS approaches the Fermi level at high doping levels, where it is not gapped like in \lscoi. Most probably this is indeed the case in bismuth compounds.\cite{kumar_characterization_2023,van_heumen_strange_2022} These parallel examples are useful cross-checks of our interpretation.

The localized hole, responsible for the pseudogap as noted above, being non-conductive, is by definition the ``non-FL'' part of the total charge active in the cuprates. Importantly, it is active, not just a charge reservoir. In fact, we argue that it plays a central role in the cuprate enigma, on two grounds. First, its universal vanishing (delocalization) on the overdoped side is concomitant with the universal vanishing of SC\revtext{, implying that the localized charge is responsible for the SC mechanism}.\cite{pelc_unusual_2019,kumar_characterization_2023,barisic_evidence_2019} Second, NMR experiments\cite{rybicki_perspective_2016} show directly that the compound-dependent charge redistribution between Cu and O with doping is related to the compound-dependent value of \Tc\revtext{:\cite{barisic_high-tc_2022} it was thus shown that the superconducting properties are directly proportional to the oxygen occupancy.\cite{rybicki_perspective_2016} The established difference between the \textbf{universal}  \emph{functional} separation of the charge between itinerant (\neff) and localized (\nloc{}) implies that localization and itinerancy are not simply determined by atomic occupation, as measured in NMR. This situation is easily understood by realizing that the localized charge \nloc{} has \emph{both} Cu and O contributions.\cite{barisic_high-tc_2022}} 

A scenario emerges in which the doped FL scatters on the localized hole, and this scattering is responsible for high-\Tc\ SC in cuprates. In this way, the localized hole introduces necessary and sufficient material-dependence into an otherwise universal FL of mobile charges. A detailed exposition of this scenario has recently been published elsewhere.\cite{barisic_high-tc_2022} Suffice it to say that the clear \emph{separation} between the FL and non-FL sector laid out here is quite different from all polaron scenarios, which rely on charge transport by these \emph{composite} electron-lattice objects, in contradiction with the observed material- and doping-independence of the FL transport parameters. It is also quite different from all scenarios which assume that the carriers in the arcs are not a FL, in contradiction with observations in all three compounds studied here with the particular purpose of elucidating that point. To repeat, the most important effective interaction in cuprates might well be the scattering of the universal FL on the localized hole, which gives rise to high-\Tc\ superconductivity. Any microscopic model of the latter must conform to the macroscopic observations presented here.

In summary, we have calculated electronic transport characteristics (the resistivity and the Hall coefficient) directly from the band structure of several cuprate materials. Combining simple FL expressions with the experimentally established pseudogapped FS's, we reproduced the observed doping evolution of the resistivity and the Hall coefficient for \hgi{}, \thalli{} and \lscoi{}. This work provides a direct link between transport coefficients and FS geometry in cuprates, showing in particular that the doping evolution of the Hall coefficient can be explained without invoking a quantum critical point or any other, even more exotic scenario. On the contrary, it is sufficient to assume that the ungapped, itinerant charge carriers are always a FL, in agreement with recent measurements of the transport and optical scattering rate. Because our approach is phenomenological, these results are observations, not hypotheses. They invite microscopic considerations on the origin of the experimental facts of Fermi arcs and universal FL scattering, which we have also briefly presented above. These are centered on the other, localized contribution to the charge-conservation equation~\eqref{eq:charcons} and its role in high-\Tc\ superconductivity. Taking the two together, we present a consistent narrative as a necessary part of any final explanation of this fascinating phenomenon.

\section*{Methods}
    \subsection{Tight-binding model parameters\label{sec:app:TB-details}}
      
\begin{figure}%
            \centering
            \includegraphics[width=0.6\textwidth]{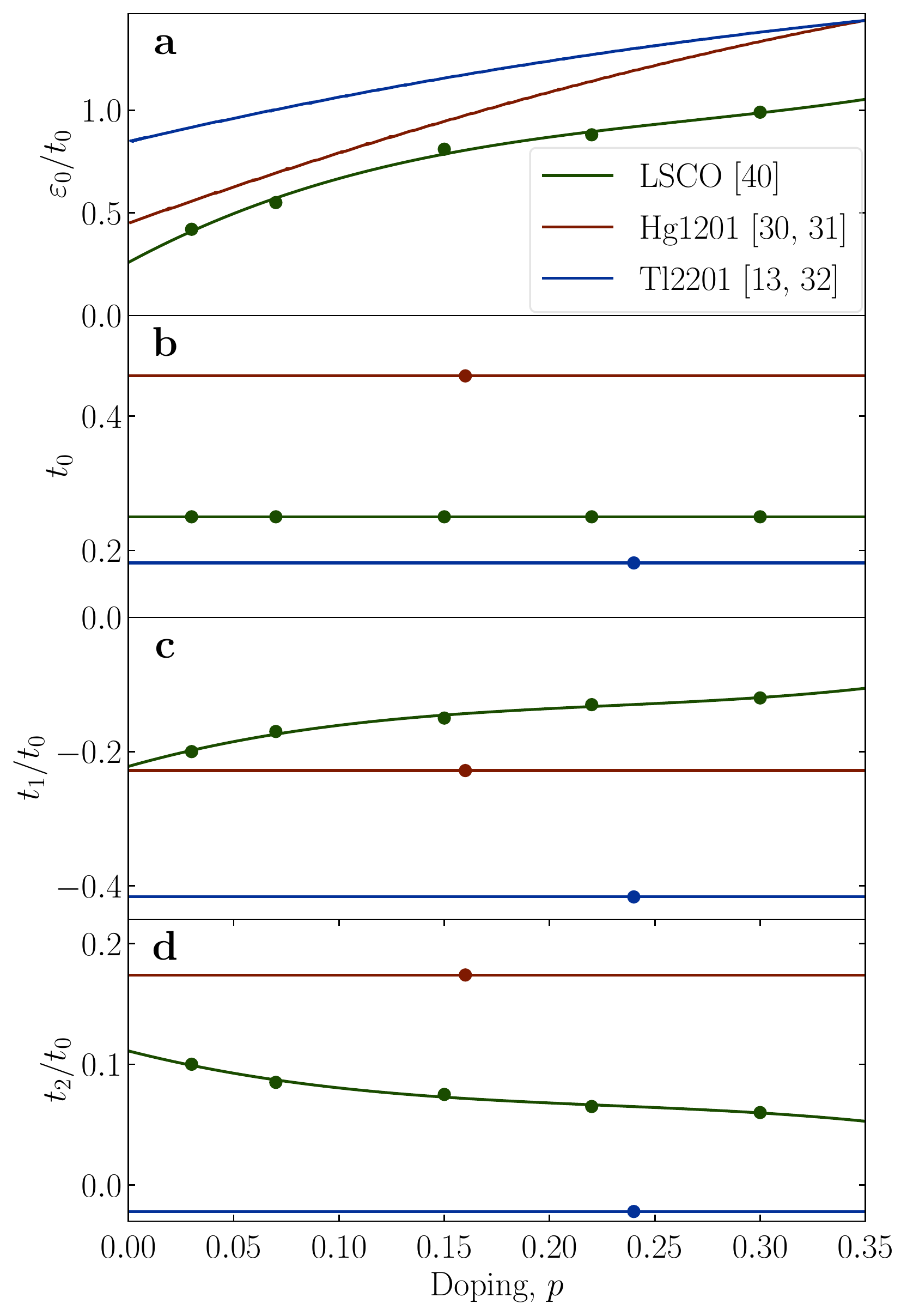}
            \caption{Doping and compound dependence of used tight-binding parameters. All parameters are given in units of [eV]. 
            Full points denote values reported in the literature according to \cref{tab:TB-parameters}, full lines are polynomial interpolations. 
            }\label{fig:TB-pars}
        \end{figure}

        Over the course of the last several decades, ARPES spectra were extensively measured and fitted with tight-binding models to parameterize the bands and the (underlying) FS's, in a number of compounds. \revtext{To eliminate a source of arbitrariness, we have strictly relied on previously published sets of parameters for each compound, even when they may be overfitted for our purposes. However, we do not expect the details of the parametrization to affect the outcomes qualitatively, as long as the experimental band structure is well described.}
        We present the tight-binding formula in a very generic form, \cref{eq:TB-model}: 
        \begin{align}
            \varepsilon_k=  \varepsilon_0 &- 2\ t_0\left[\cos{\left(k_xa\right)} +\cos{\left(k_ya\right)}\right] \nonumber \\
                           &- 4\ t_1\cos{\left(k_xa\right)}\cos{\left(k_ya\right)} \nonumber \\
                           &- 2\ t_2\left[\cos{\left(2k_xa\right)} +\cos{\left(2k_ya\right)}\right] \nonumber \\
                           &- 4\ t_3\left[\cos{\left(2k_xa\right)}\cos{\left(k_ya\right)}  +\cos{\left(k_xa\right)}\cos{\left(2k_ya\right)}\right] \nonumber \\
                           &+ 0.5\ t_4\cos{\left(2k_xa\right)}\cos{\left(2k_ya\right)}    \label{eq:TB-model}
        \end{align}
        \noindent where the tight-binding parameters for compounds studied in this work (\hgi{} \cite{das_q0_2012,vishik_angle-resolved_2014},  \thalli{} \cite{plate_fermi_2005,peets_tl2ba2cuo6_2007} and  \lscoi{} \cite{yoshida_low-energy_2007}) are given in \cref{tab:TB-parameters} and visualized in \cref{fig:TB-pars}. Because the naming convention is standardized\revtext{, some parameters are zero for \hgi{} and \lscoi{}, simplifying Eq.~\eqref{eq:TB-model} in those cases. }
                
        \begin{table}
            \centering
            \caption{Tight binding parameters for the FS models in use. 
            In the case of Hg1201 and Tl2201, the function for $\varepsilon_0$ is determined to satisfy \cref{eq:rigid_band_shift}, while all other parameters are held fixed, as in Ref.~\citeonline{vishik_angle-resolved_2014}. 
            In contrast, for \lscoi\ a broad range of doping dependent parameters exist, see \cref{fig:TB-pars}. 
            All numerical parameters are given in units of [eV]. 
            Parameters described as doping dependent functions are displayed in \cref{fig:TB-pars}.
            }\label{tab:TB-parameters}%
            \begin{tabular}{cccc}
            \toprule
            Parameter & Hg1201 \cite{das_q0_2012,vishik_angle-resolved_2014}  
                      & Tl2201 \cite{plate_fermi_2005,peets_tl2ba2cuo6_2007} 
                      & \lscoi\ \cite{yoshida_low-energy_2007} \\
            \midrule
            $\varepsilon_0$ & $\varepsilon_0 = \varepsilon_0(p)$ & $\varepsilon_0 = \varepsilon_0(p)$ & $\varepsilon_0 = \varepsilon_0(p)$  \\ 
            $t_0$           & 0.46                               & 0.18125                            & 0.25                                \\ 
            $t_1$           & -0.105                             & -0.0755                            & $t_1 = t_1(p)$                      \\ 
            $t_2$           & 0.08                               & -0.003975                          & $-0.5t_1$                           \\ 
            $t_3$           & -0.02                              & -0.0100625                         & 0                                   \\ \addlinespace 
            $t_4$           & 0                                  & 0.0068                             & 0                                   \\ \bottomrule 
            \end{tabular}
        \end{table}            

        \begin{table}
            \centering
            \caption{Parameters of the gap distribution used to determine \neff{}, according to the approach described in Ref.~\citeonline{pelc_unusual_2019}. }
            \label{tab:gap-parameters}
            \begin{tabular}{cccc}
                \toprule
                Parameter & \hgi{}  & \thalli{}  & \lscoi\  \\  
                \midrule
                $\Delta_0$       & 4000   & 3700   & 3900 \\ 
                $\delta$       & 600    & 700    & 800  \\ 
                $p_c$       & 0.2    & 0.22   & 0.22 \\ 
                $\alpha$        & 2      & 2      & 2   \\ 
                \bottomrule
            \end{tabular}
        \end{table}

        In the case of \lscoi{}, high-quality ARPES data exists for a range of doping levels, therefore it is possible to extract the evolution of the FS with doping directly. 
        For \hgi{} and \thalli{}, the number of doping levels on which ARPES studies have been performed is much more limited (one doping level each). 
        Therefore, the doping dependence for these materials is introduced by a rigid band shift, respecting Luttinger's sum rule in the underlying FS:
        \begin{equation}
            1+p = 2 \frac{A^{u}_{FS}}{A_{BZ}}\label{eq:rigid_band_shift}
        \end{equation}
        
        \noindent where $A^u_{FS}$ denotes the surface area enclosed by the underlying FS in k-space, while $A_{BZ}$ denotes the area of the first Brillouin zone. 

    \revtext{The published parametrization we used is only optimized to follow the main properties of the band intersecting the Fermi level and particularly the evolution of the Fermi surface. The kink at $\sim 50$~meV, presumably due to strong correlations in the Cu 3d orbital, is not taken into consideration. We believe that the kink comes from the Fermi liquid probing the still-localized 3d orbital. Furthermore, it is often missed that another channel is open for transport that does not involve the 3d orbital. This is the Cu 4s orbital via the second-order 2p-4s-2p hopping,\cite{Pavarini01} as argued in previous publications.\cite{barisic_high-tc_2022} Because the 4s orbital is large, larger even than the 2p orbital, the 4s-2p overlap $t_{ps}$ can be very large. Notably, the need for unreasonably large effective values of $t_{pp}\sim 1$~eV, when fitting ARPES in the Emery three-band model without the 4s orbital, as well as the universality of transport properties, is direct indication that the electrons are really taking advantage of the 4s orbital.}
    
\revtext{Experiments (e.g., ARPES) show very different behavior of nodal and antinodal parts of the Fermi
surface, which is the k-space signature of the separation into itinerant and localized electrons. For the system to be stable, the chemical potential must be the same for both sectors. Because the itinerant sector is a Fermi liquid, any non-Fermi-liquid behavior of the chemical potential must be ascribed to the localized part, the latter (obviously) not being a Fermi liquid. However, we point out once more that the underlying Fermi surface always contains $1+p$ states, which is in agreement with the Luttinger sum rule.}

    \subsection{Details of the calculation procedure}

        \subsubsection{Circular FS (parabolic band)  -- ungapped \label{sec:app:calc_circular_analytical}}
            
            Using the general expressions \cref{eq:sigma_xx,eq:sigma_xy}, it is instructive to derive $\sigma_{ij}$ for the particularly simple case of a circular FS and isotropic group velocity, $v_F=\hbar^{-1}|\partial\varepsilon/\partial k_\perp|=\hbar k_F/m^*$, where $m^*$ is the effective mass.
            
            \begin{equation}
                \sigma_{xx}=\frac{e^2\tau n}{m^*}\label{sigmaxx_drude}
            \end{equation}
            
            \noindent where the concentration of charge carriers $n$ may be expressed in terms of the ratio of the area of the occupied part of the Brillouin zone and the total area of the Brillouin zone,
            
            \begin{equation}
            n=(2s+1)\;n_0\frac{k_F^2\pi}{\Gamma_{2D}}\;.\label{FSurRat}
            \end{equation}
            
            \noindent The most simple Drude form for $\sigma_{xx}$ in \cref{sigmaxx_drude}, which is obtained from the general expression in \cref{eq:sigma_xx}, is a consequence of the particular circular-shaped form of the FS and the fact that the velocity is in the same direction and proportional to $k_F$ along the FS, $v_F\propto k_F$.  
            
            With the circular FS and the constant velocity $\abs{v_F}$, the nondiagonal part of the conductivity tensor is given by 
            
            \begin{align}
                \sigma_{xy} &=(\omega_c\tau)\;\frac{e^2\tau}{m^*}\; (2s+1)\;n_0\frac{k_F^2\pi}{\Gamma_{2D}}\\
                            &=(\omega_c\tau)\;\sigma_{xx}
            \end{align}
            
            \noindent with $\sigma_{xx}$ given by \cref{sigmaxx_drude} and $\omega_c$ the cyclotron frequency, $\omega_c=eB/m^*$. For a parabolic band, the effective mass $m^*$, characterized by the second derivative of the dispersion at the bottom of the band, is the only model parameter that defines the dispersion at the FS for any doping. However, this simplicity is lost for any more complicated band structure.

        \subsubsection{Circular FS (parabolic band) -- gapped}\label{sec:parabolic_band}

            Assuming a circular FS that does not intersect the zone boundaries, following Luttinger's theorem one obtains
            
            \[k_F=\left(\frac{1+x}{2\pi}\right)^{\frac{1}{2}}\frac{\pi}{a}\;.\]
            
            \noindent Introducing $0\leq p(x)\leq1$, as a parameter that defines the ungapped part of the FS, the concentration of itinerant charges in \cref{FSurRat} takes a particularly simple form,
            
            \begin{equation}
                n=(2s+1)\;n_0\;\frac{1+x}{2}\;p(x)\;.
            \end{equation}

\begin{figure}%
                \centering
                \includegraphics[width=1\textwidth]{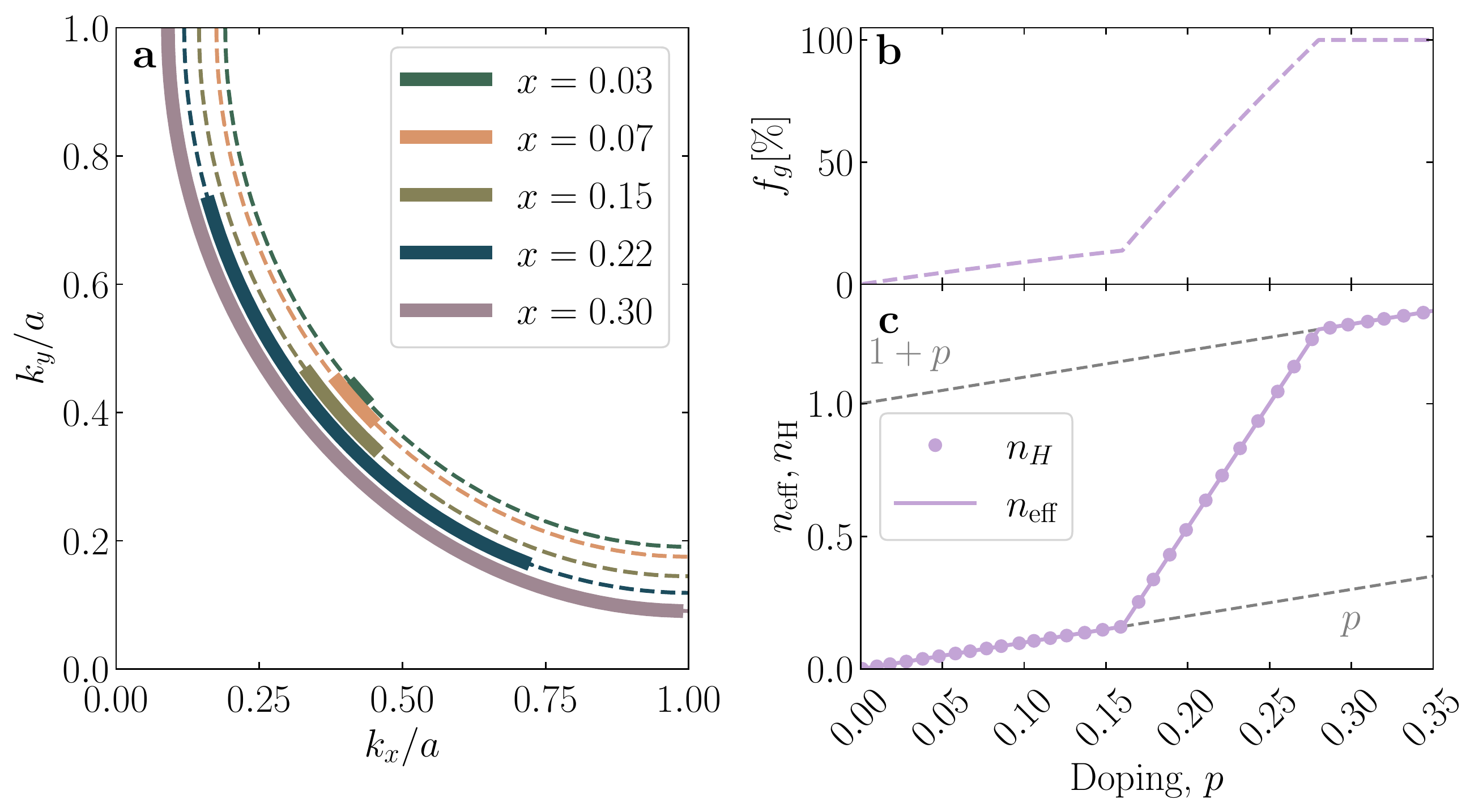}
                \caption{FS and Hall-coefficient of an ideal parabolic band. 
                In \textbf{a}, dashed lines correspond to the underlying FS, while the arcs appear as full lines.
                The fraction of ungapped (i.e. "active") states $f_g$ is displayed in \textbf{b}, and the calculated density of charge carriers \neff{}(line) and \nH\ at selected doping levels (points) in \textbf{c}. 
                In this case, the crossover from \ptop{} is modeled between $p=0.16$ and $p=0.28$.
                }\label{fig:Summary_parabola}
            \end{figure}

            Assuming a parabolic band, exhibiting a circular FS and an isotropic group velocity, we consider a doping-dependent gapping mechanism that resembles the situation in cuprates.  We chose $f_g$ in a way that \neff{} first evolves exactly as $p$ (for $p < 0.16$ ) to increase more steeply to $1+p$ at $p = 0.28$. From \cref{fig:Summary_parabola}, it is obvious that this results in a 1:1 correspondence between \nH\ and \neff{}.

    \subsection{Comparison with experimental data: Resistivity}  \label{sec:methods:resistivity}

        Our calculations are performed in the low-temperature limit. To compare calculation results with experimental data collected at finite temperatures, we introduce a temperature independent variable following the arguments discussed  below.        
        A general expression (Taylor expansion)  for the resistivity is: 
         \begin{align}\label{eq:resistivity}
            \rho &= A_0 + A_1 T + A_2 T^2    
        \end{align}        
         \noindent where $A_0$ is associated with sample-dependent impurity scattering, $A_1$ that appears in the crossover/strange metal regime we attribute to a change in the carrier density due to the delocalization process discussed in the main text, while $A_2$ is the Fermi-liquid term (associated with \neff{($T = 0$ K)} charges). Thus, the coefficient $A_2$ is of our main interest. Specifically, we use $A_{2,\Box}$, as the resistivity per CuO$_2$ was demonstrated to be universal across multiple cuprate families.\cite{barisic_universal_2013} 

        We extract the scattering time $\tau$, presumably related to the Umklapp process,\cite{tabis_arc--pocket_2021} from the measured universal Hall-mobility:\cite{barisic_evidence_2019}
        \begin{align}
            \mu_H &= \frac{e\tau}{m^{*}} \\ 
            \mu_H^{-1} &= C_0 + C_2 T^2 
        \end{align}
        \noindent As in the case of the resistivity, the constant term $C_0$ is a contribution related to the impurities.\cite{chien_effect_1991} 
        We approximate the effective mass with a constant $m^{*}\sim 3.5m_e$ (\cref{eq:tau-from-muH}), again because the universality of the Hall mobility (\cref{fig:intro_C2}) implies it, where the exact value was determined from quantum oscillations in overdoped \thalli{}. Notably, we do expect some compound-dependence of the effective mass but such corrections are not essential in the context of the present calculations.

        \begin{align}
            \tau &= \frac{m^{\ast}}{e} \cdot \mu_H \nonumber \\
                 &\simeq \frac{3.5 m_e}{e} \cdot \frac{1}{C_2 T^2} \label{eq:tau-from-muH}
        \end{align}        

        Combining the two $T^2$-like behaviors, we arrive at a temperature independent parameter $\tau\rho_{\Box}$: 
        
        \begin{align}
            \tau \rho_{\Box} &= \frac{A_{2,\Box} T^2}{C_2 T^2} \frac{m^{\ast}}{e} \nonumber \\
                      &= \frac{A_{2,\Box}}{C_2} \frac{m^{\ast}}{e}
        \end{align}

\section*{Data Availability}
The datasets generated and/or analysed during the current study are available from the corresponding authors upon request.

\section*{Code Availability}
        The code to reproduce the presented results is available from the corresponding authors upon request.


\section*{Acknowledgments}
The work at the TU Wien was supported by the European Research Council (ERC Consolidator Grant No. 725521), while the work at the University of Zagreb was supported by project CeNIKS co-financed by the Croatian Government and the European Union through the European Regional Development Fund-Competitiveness and Cohesion Operational Programme (Grant No. KK.01.1.1.02.0013). The work at AGH University of Krakow was supported by the National Science Centre, Poland, Grant No. OPUS: UMO-2021/41/B/ST3/03454, the Polish National Agency for Academic Exchange under “Polish Returns 2019” Programme: PPN/PPO/2019/1/00014, and the subsidy of the Ministry of Science and Higher Education of Poland. M.A.G. was partly supported by program „Excellence Initiative – Research University” for AGH University of Krakow. O.S.B. acknowledges the support by the QuantiXLie Center of Excellence, a project co-financed by the Croatian Government and European Union through the European Regional Development Fund - the Competitiveness and Cohesion Operational Programme (Grant KK.01.1.1.01.0004).

\section*{Author contributions statement}

N.B. conceived the research, B.K.K and O.S.B. performed the analysis and computations. W.T. and M.A.G. verified the results and methods. B.K.K., W.T., D.K.S. and N.B. wrote the manuscript with input from all authors.

\section*{Additional information}

The authors declare no competing interests. 

\end{document}